\documentclass[12pt]{article}
\usepackage{graphics}
\usepackage{epsfig}

\begin{document}

\setlength{\textheight}{240mm}
\voffset=-15mm
\baselineskip=20pt plus 2pt
\renewcommand{\arraystretch}{1.6}

\begin{center}

{\large \bf  The Energy of Regular Black Hole in General Relativity Coupled
to Nonlinear Electrodynamics}\\
\vspace{5mm}
\vspace{5mm}
I-Ching Yang$^{\dag}$ \footnote{E-mail:icyang@nttu.edu.tw},
Chi-Long Lin$^{\ddag}$ 
and $^{\S}$Irina Radinschi \footnote{E-mail:radinschi@yahoo.com}

$^{\dag}$Department of Natural Science Education \\
and Systematic and Theoretical Science Research Group, \\
National Taitung University, Taitung, Taiwan 950,\\
$^{\ddag}$The National Museum of Natural Science, \\
Taichung, Taiwan 403, Republic of China, \\
and $^{\S}$Department of Physics, "Gh. Asachi" Technical University, \\
Iasi, 700050, Romania

\end{center}
\vspace{5mm}

\begin{center}
{\bf ABSTRACT}
\end{center}
According to the Einstein, Weinberg, and M{\o}ller energy-momentum complexes, we evaluate the 
energy distribution of the singularity-free solution of the Einstein field equations coupled to a 
suitable nonlinear electrodynamics suggested by Ay\'{o}n-Beato and Garc\'{i}a.  The results show 
that the energy associated with the definitions of Einstein and Weinberg are the same, but M{\o}ller 
not.  Using the power series expansion, we find out that the first two terms in the expression are 
the same as the energy distributions of the Reissner-Nordstr\"{o}m solution, and the third term 
could be used to survey the factualness between numerous solutions of the Einstein field eqautions 
coupled to a nonlinear electrodynamics.

\vspace{2mm}

\vspace{5mm}
\noindent

\newpage

\section{Introduction}

In general relativity most of the solutions of the Einstein equations exhibit the same important 
property, which is the existence of singularities~\cite{1}.  The space-time at the center of a 
black hole presents an infinite curvature and under the action of infinite gravity the matter is 
crushed to infinite density.  The problem that arises is the impossibility of the laws of physics 
to hold at the singularity.  Important theories, like the Brans-Dicke theory and the Einstein-Cartan 
theory, couldn't yield a satisfactory solution for avoiding the existence of the singularity in their 
solutions.  To avoid the problem of singularity, some regular models have been proposed~\cite{2}.  
These solutions are known as ``Bardeen black holes"~\cite{3}, as Bardeen elaborated for the first 
time an interesting regular black hole model.  These models are not exact solutions to Einstein 
equations, because no known physical source is associated with any of them and the search of the 
best candidate which can produce singularity-free solutions has continued.  A way to solve this 
problem is to find more general gravity theories avoiding the existence of singularities.  String 
theory~\cite{4} produces singularity-free solutions, even at the classical level, due to its 
intrinsic non-locality.  Other examples are given by domain wall solutions with horizons but without 
singularities in ${\cal N} =1$ supergravity (cf.~\cite{5}, and references therein) and exact conformal 
field theory~\cite{6}.  Ay\'{o}n-Beato and Garc\'{i}a~\cite{7,8} show that in the framework of the 
standard general relativity it is possible to generate singularity-free solutions of the Einstein 
field equations coupled to a suitable nonlinear electrodynamics, which in the weak field approximation 
becomes the usual linear Maxwell theory.  The solutions are given by the line element 
\begin{equation}
ds^2 =A(r) dt^2 - A(r)^{-1} dr^2 -r^2 d\Omega .
\end{equation}
In this article, we evaluate the energy distributions of above regular black hole soution by using 
the Einstein, Weinberg, and M{\o}ller energy-momentum complexes.  Through the paper we use geometrized 
units ($G=1$, $c=1$), and follow the convention that Latin indices run from 1 to 3 and Greek indices run
from 0 to 3.

\section{Three Pseudotensoeial Prescriptions for Two Regular Black Hole Solutions}

Let us consider about the first regular black hole solution is presented by Ay\'{o}n-Beato and 
Garc\'{i}a~\cite{7} in 1998 with 
\begin{equation}
A(r) = 1 - \frac{2mr^2}{(r^2+q^2)^{3/2}} + \frac{q^2 r^2}{(r^2+q^2)^2} .
\end{equation}
This solution asymptotically behaves as the Reissner-Nordstr\"{o}m solution and the parameters $m$ and 
$q$ represents the mass and the electric charge.   At the outset, the energy component of the Einstein 
energy-momentum complex~\cite{9} is given by
\begin{equation}
E_{\rm Einstein} = \frac{1}{16\pi} \int \frac{\partial H_0^{\;\;0l}}{\partial x^l} d^3x ,
\end{equation}
where 
\begin{equation}
H_0^{\;\;0l} = \frac{g_{00}}{\sqrt{-g}} \frac{\partial}{\partial x^m} \left[ (-g) g^{00} g^{lm} \right] .
\end{equation}
The energy component of the Einstein energy-momentum complex is most conveniently calculated in 
the quasi-Cartesian coordinates $(t,x,y,z)$.  In these coordinates, the line element Eq.(1) reads
\begin{equation}
ds^2 = A dt^2 - (dx^2 + dy^2 + dz^2) - \frac{A^{-1}-1}{r^2} (x dx + y dy + z dz)^2 .
\end{equation}
Then, the required nonvanishing components of the Einstein energy-momentum complex $H_0^{\;\;0l}$
are easily shown in spherical coordinates to be
\begin{equation}
H_0^{\;\;0r} = \frac{2 \kappa}{r} \hat{r} - \frac{1}{A} \hat{r} (\hat{r} \cdot \nabla A) 
+\frac{1}{A} \nabla A  , 
\end{equation}
where $\kappa = 1 - A$.  Applying the Gauss theorem we obtain
\begin{equation}
E_{\rm Einstein} = \frac{1}{16\pi} \oint H_0^{\;\;0r} \cdot \hat{r} r^2 d\Omega  ,
\end{equation}
and the integral being taken over a sphere of radius $r$, with the outward normal $\hat{r}$ and 
the differential solid angle $d\Omega$.  The Einstein energy complex within radius $r$ reads
\begin{equation}
E_{\rm Einstein} = \frac{r}{2} (1 - A) = m (1+\frac{q^2}{r^2})^{-3/2} -\frac{q^2}{2r} 
(1+\frac{q^2}{r^2})^{-2} \equiv E_1 (r).
\end{equation}
Next, the Weinberg energy-momentum complex~\cite{10} is considered as
\begin{equation}
\tau ^{\nu \lambda} = \frac{\partial}{\partial x^{\rho}} Q^{\rho \nu \lambda} ,
\end{equation}
with superpotential
\begin{equation}
Q^{\rho \nu \lambda} = \frac{\partial h^{\mu}_{\mu}}{\partial x_{\rho}} 
\eta^{\nu \lambda} - \frac{\partial h^{\mu}_{\mu}}{\partial x_{\nu}} 
\eta^{\rho \lambda} + \frac{\partial h^{\mu \nu}}{\partial x^{\mu}} 
\eta^{\rho \lambda} - \frac{\partial h^{\mu \rho}}{\partial x^{\mu}} 
\eta^{\nu \lambda} - \frac{\partial h^{\nu \lambda}}{\partial x_{\rho}} 
+ \frac{\partial h^{\rho \lambda}}{\partial x_{\nu}}  ,
\end{equation}
where $\eta_{\mu \nu}$ is the Minkowski metric and $h_{\mu \nu} = g_{\mu \nu} 
- \eta_{\mu \nu}$.  We adopt the convenient convention that the indices on 
$h_{\mu \nu}$ and $ \partial / \partial x^{\lambda} $ are raised and lowered 
with $ \eta $.  The energy-momentum of the Weinberg energy-momentum complex 
is most conveniently calculated in the quasi-Cartesian coordinates $(t,x,y,z)$, 
and is given by
\begin{equation}
P^{\lambda} = \frac{1}{16\pi} \int \frac{\partial Q^{i0\lambda}}
{\partial x^i} d^3x .
\end{equation}
The reguired nonvanshing components $Q^{i00}$ of the Weinberg energy complex are easily
shown in spherical coordinates to be
\begin{equation}
Q^{i00} = \frac{\eta}{r} \hat{r} +\frac{\hat{r}}{2} (\hat{r} \cdot \nabla \eta) 
-\frac{1}{2} \nabla \eta ,
\end{equation}
where $\eta = A^{-1} -1$.  Applying the Gauss theorem, hence, the energy within radius $r$ 
obtained from the Weinberg complex is
\begin{equation}
E_{\rm Weinberg} = P^0 = \frac{1}{16\pi} \oint Q^{i00} n_i r^2 d\Omega =\frac{r}{2} \eta  .
\end{equation}
The energy component of the covariant energy-momentum four vector of the Weinberg 
energy-momentum complex is 
\begin{equation}
E_{\rm Weinberg}^{covariant} = g_{00} E_{\rm Weinberg} = \frac{\kappa r}{2} 
= E_{\rm Einstein} .
\end{equation}

Subsequently, in the M{\o}ller prescription the energy-momentum complex~\cite{11} which is given by 
\begin{equation}
\Theta _\nu ^{\;\;\mu }=\frac 1{8\pi }\frac{\partial \chi _\nu ^{\;\;\mu
\sigma }}{\partial x^\sigma },
\end{equation}
where the M{\o}ller superpotential 
\begin{equation}
\chi _\nu ^{\;\;\mu \sigma }=\sqrt{-g}\left( \frac{\partial g_{\nu \alpha }}{%
\partial x^\beta }-\frac{\partial g_{\nu \beta }}{\partial x^\alpha }\right)
g^{\mu \beta }g^{\sigma \alpha }  
\end{equation}
are quantities antisymmetric in the indices $\mu $, $\sigma $. According to the definition of the 
M{\o}ller energy-momentum complex, the expression for energy is given as 
\begin{equation}
E_{\rm M{\o}ller} = \frac {1}{8\pi} \int \frac{\partial \chi _0^{\;\;0k}}{\partial x^k}
d^3 x  .
\end{equation}
Notice, that the only nonvanshing component of the M{\o}ller energy-momentum complex is 
\begin{equation}
\chi _0^{\;\;0k} = r^2 \sin \theta \frac{dA}{dr},
\end{equation}
and the M{\o}ller energy within radius $r$ is 
\begin{equation}
E_{\rm M{\o}ller} = \frac{r^2}{2} \frac{dA}{dr} = m (1+\frac{q^2}{r^2})^{-5/2} 
(1-\frac{2q^2}{r^2}) - \frac{q^2}{r} (1+\frac{q^2}{r^2})^{-3} (1-\frac{q^2}{r^2})  
\equiv E_2 (r) .
\end{equation}

Once more, Ay\'{o}n-Beato and Garc\'{i}a show another new regular black hole solution 
in 1999 ~\cite{8} with 
\begin{equation}
A(r)= 1 - \frac{2mr^2 e^{-q^2 /2mr}}{(r^2+q^2)^{3/2}}  ,
\end{equation}
and the parameters $m$ and $q$ are associated with mass and charge respectively.  For this
black hole solution, the Einstein energy complex is 
\begin{equation}
E_{\rm Einstein} = m (1+\frac{q^2}{r^2})^{-3/2} e^{-q^2 /2mr}  
\equiv E_3 (r)  ,
\end{equation}
and the M{\o}ller energy complex obtained by Radinschi~\cite{12} is
\begin{equation}
E_{\rm M{\o}ller} = \left[ m (1+\frac{q^2}{r^2})^{-5/2} (1-\frac{2q^2}{r^2}) - 
\frac{q^2}{2r} (1+\frac{q^2}{r^2})^{-3/2} \right] e^{-q^2 /2mr} \equiv E_4 (r).
\end{equation}
On the other hand, using the power series expansion, the energy distribution of Einstein 
energy-momentum complex would become 
\begin{equation}
E_1 (r) = E_{\rm Tod} - \frac{3m q^2}{2 r^2} + \frac{q^4}{r^3} 
+\frac{15m q^4}{8 r^4} -\frac{3 q^6}{2 r^5} + {\cal O} (\frac{1}{r^6})  ,
\end{equation}
and 
\begin{eqnarray}
E_3 (r) & = & E_{\rm Tod} + (\frac{1}{8m} - \frac{3m}{2q^2}) \frac{q^4}{r^2}  
- (\frac{1}{48m^2} - \frac{3}{4q^2}) \frac{q^6}{r^3} \nonumber \\
& & + (\frac{1}{384m^3} -\frac{3}{16mq^2} +\frac{15m}{8q^4}) \frac{q^8}{r^4} \nonumber \\
& & - (\frac{1}{3840m^4} -\frac{1}{32m^2q^2} +\frac{15}{16q^4}) \frac{q^{10}}{r^5} 
+ {\cal O} (\frac{1}{r^6})  ,
\end{eqnarray}
where the term $E_{\rm Tod}$ represents the energy of the Reissner-Nordstr\"{o}m 
solution that corresponds to the Penrose quasi-local mass definition (evaluated by 
Tod~\cite{13}).  Also, the energy distribution of M{\o}ller energy-momentum complex would 
become
\begin{equation}
E_2 (r) = E_{\rm Komar} - \frac{9m q^2}{2r^2} + \frac{4 q^4}{r^3} 
+\frac{75m q^4}{8 r^4} -\frac{9 q^6}{ r^5} + {\cal O} (\frac{1}{r^6})  ,
\end{equation}
and 
\begin{eqnarray}
E_4 (r) & = & E_{\rm Komar} + (\frac{3}{8m} - \frac{9m}{2q^2}) 
\frac{q^4}{r^2} - (\frac{1}{12m^2} - \frac{3}{q^2}) \frac{q^6}{r^3} \nonumber \\
& & + (\frac{5}{384m^3} -\frac{15}{16mq^2} +\frac{75m}{8q^4}) \frac{q^8}{r^4} \nonumber \\
& & - (\frac{1}{640m^4} -\frac{3}{16m^2q^2} +\frac{45}{8q^4}) \frac{q^{10}}{r^5} 
+ {\cal O} (\frac{1}{r^6})  ,
\end{eqnarray}
where the term $E_{\rm Komar}$ agrees with the energy of the Reissner-Nordstr\"{o}m 
solution in the Komar prescription~\cite{14}.  The first two terms in all expressions 
are the same as the energy distribution of the Reissner-Nordstr\"{o}m solution, and the 
energy distributions asymptotically behaves as the Reissner-Nordstr\"{o}m solution.

\section{Conclusion}

One of the most important themes of general relativity, the energy-momentum localization has 
not yield a satisfactory solving. At international level, considerable efforts have been made 
to find a generally accepted expression for the energy-momentum density.  Many scientist worked 
at this issue, using different definitions, like superenergy tensors~\cite{15}, quasi-local mass 
definitions~\cite{16,17,18,19}, pseudotensorial prescriptions, and here we notice the 
energy-momentum complexes of Einstein~\cite{9}, Landau-Lifshitz~\cite{20}, Papapetrou~\cite{21}, 
Weinberg~\cite{10} (ELLPW), Bergmann-Thomson~\cite{22}, Qadir-Sharif~\cite{23} and 
M{\o}ller~\cite{12}, and teleparallel gravity theory~\cite{24}.  The pseudotensorial definitions, 
except the M{\o}ller energy-momentum complex, imply to performed the calculations using non-covariant, 
coordinate dependent expressions, and yields acceptable results only in the case of quasi-Cartesian 
coordinates.  In the recent years the problem of the usefulness of energy-momentum complexes became 
a re-opened issue, and many interesting results were obtained~\cite{25}, which demonstrated that these 
definitions are powerful concepts for energy-momentum localization in general relativity. The (ELLPW), 
Bergmann-Thompson and M{\o}ller prescriptions yield meaningful results in the case of 2 and 3 dimensional 
space-times~\cite{26}. We also point out another interesting issue, the similarity of results which were 
obtained for many geometries using both energy-momentum definitions and the teleparallel gravity 
theory~\cite{24}.  Whereas Chang, Nester and Chen~\cite{16} demonstrated in their studies that the 
energy-momentum complexes can be considered actually quasi-local and legitimate expressions for the 
energy-momentum.  Their conclusion is that there exist a direct connection between energy-momentum 
complexes and quasi-local expressions.  Furthermore, the new idea of quasi-local approach for 
energy-momentum complexes~\cite{16,17} is the subject of interesting studies and a large class of new 
pseudotensors connected to the positivity in small regions have been elaborated~\cite{18}.  The 
quasi-local quantities can be associated with a closed 2-surface~\cite{17}.  The quasi-local quantities 
for finite regions are determined by the Hamiltonian boundary term and the special quasi-local 
energy-momentum boundary term expressions are connected to physically distinct and geometrically clear 
boundary conditions~\cite{19}.  Our paper is focused 
on the evaluation of the energy distribution for two regular black hole solutions in general relativity 
coupled to nonlinear electrodynamics given by Ay\'{o}n-Beato and Garc\'{i}a~\cite{7,8}.

In this paper,  according to the definations of the energy-momentum 
pseudotensor of Einstein, Weinberg, and M{\o}ller, we evaluate the energy distributions of the 
singularity-free solution, which is obtained by Ay\'{o}n-Beato and Garc\'{i}a, from the Einstein 
equations coupling to a nonlinear electrodynamics.  Also, the Einstein and Weinberg energy-momentum 
complex have the same results, and the M{\o}ller energy-momentum complex gives a different result.  
In the limit of a vanishing charge $q$, we obtain in the Einstein, Weinberg and M{\o}ller prescription 
the same result, the energy is the mass of the black hole, and this is the same as the expression 
for the energy of the Schwarzschild solution.  Here, Vagenas~\cite{27} hypothesizes that there 
is a relation 
\begin{equation}
\alpha_n^{\rm (Einstein)} = \frac{1}{n+1} \alpha_n^{\rm (M{\o}ller)}  ,
\end{equation}
between the cofficients of the expression for energy in the Einstein prescription
\begin{equation}
E_{\rm Einstein} = \sum_{n=0}^{\infty} \alpha_n^{\rm (Einstein)} r^{-n}
\end{equation}
and in the M{\o}ller prescription 
\begin{equation}
E_{\rm M{\o}ller} = \sum_{n=0}^{\infty} \alpha_n^{\rm (M{\o}ller)} r^{-n}  .
\end{equation}
To compare with Eq.(23), Eq.(24), Eq.(25) and Eq.(26), our results support Vagenas hypothesize.  
Although, the relation can be understood the product of the formula~\cite{28}
\begin{equation}
E_{\rm M{\o}ller} = E_{\rm Einstein} - r \frac{d E_{\rm Einstein}}{dr}  ,
\end{equation}
which will be derived from Eq.(8) and Eq.(19).  Furthermore, we make a comparision with 
the expression for energy of Einstein and M{\o}ller energy-momentum complexes obtained by 
Radinschi~\cite{29} 
\begin{equation}
\bar{E}_{\rm Einstein} = \bar{E}_{\rm Weinberg} = E_{\rm Tod} + \frac{q^6}{24 m^2 r^3} - 
\frac{q^{10}}{240 m^4 r^5} + {\cal O} (\frac{1}{r^6}) 
\end{equation}
and
\begin{equation}
\bar{E}_{\rm M{\o}ller} = E_{\rm Komar} + \frac{q^6}{6 m^2 r^3} - \frac{q^{10}}{40 m^4 r^5}
+{\cal O} (\frac{1}{r^6})  . 
\end{equation}
These results are evaluated for another regular black hole solution which is also suggested 
by Ay\'{o}n-Beato and Garc\'{i}a~\cite{30}.  Notice to the difference in the third term 
between these two solutions, it could be used to survey the factualness of solutions.

\end{document}